\title{Generalizing the Kodama State II: Properties and Physical Interpretation}
\author{
        Andrew Randono \\
               Center for Relativity, Department of Physics\\
        University of Texas at Austin \\
        Austin, TX 78712\\
                  email: arandono@physics.utexas.edu
}

\documentclass[12pt]{article}
\usepackage{amssymb}
\newcommand{\beq}{\begin{equation}}
\newcommand{\eeq}{\end{equation}}
\newcommand{\beqa}{\begin{eqnarray}}
\newcommand{\eeqa}{\end{eqnarray}}
\newcommand{\w}{\wedge}

\newcommand{\nn}{\nonumber}

\begin{document}
\maketitle
\bibliographystyle{utphys}
\begin{abstract}
In this second part of a two paper series we discuss the properties and physical interpretation of the generalized
Kodama states. We first show that the states are the three dimensional boundary degrees of freedom of two familiar 4-dimensional
topological invariants: the second Chern class and the Euler class. Using this, we show that the states have the
familiar interpretation as WKB states, in this case corresponding
not only to de Sitter space, but also to first order perturbations therein. In an appropriate spatial
topology, the de Sitter solution has pure Chern-Simons functional form, and is the unique state in the class that is
identically diffeomorphism and $SU(2)$ gauge invariant. The q-deformed loop transform of this state yields evidence of
a cosmological horizon when the deformation parameter is a root of untiy. We then discuss the behavior of the states
under discrete symmetries, showing that the states violate $P$ and $T$ due to the presence of the Immirzi parameter,
but they are $CPT$ invariant. We conclude with an interesting connection between the physical inner product and the
Macdowell Mansouri formulation of gravity.
\end{abstract}
\section{Physical Interpretation}
 In the first paper of this set, we showed that the generalized Kodama states
share many properties in common with the ordinary single particle momentum eigenstates, properties which we exploited in
the construction of the states. In this section, we will show that the analogy can be extended to the physical
interpretation
as well---in common with the momentum eigenstates, the generalized Kodama states are WKB states in addition to
being exact quantum states. We will suggest that, in contrast to the original Kodama state, the set of generalized states
includes not only a WKB analogue of de-Sitter space, but also small vacuum perturbations to de Sitter space. We
suggest that this may explain why there appears to be a large class of states. 

Let us first briefly review the WKB construction of the momentum eigenstates. We recall that in the WKB approximation,
the wave function is split into amplitude and phase, $\Psi=\rho e^{i \Theta /\hbar}$. This splits the Schr\"{o}dinger
equation into two pieces---one piece expresses the conservation of the probability current density, and the other piece
is the Hamilton-Jacobi equation for $\Theta$ plus a small correction proportional to $\hbar$ that is interpreted as a
quantum potential. The WKB approximation consists of solving the Schr\"{o}dinger equation in successive powers of
$\hbar$. Since the quantum potential is proportional to $\hbar$, to lowest order in $\hbar$, the wave function can be
approximated by $\Psi \simeq e^{i S_{0}/\hbar}$ where $S_{0}$ is a solution to the Hamilton-Jacobi equation. A preferred
solution to the Hamilton-Jacobi equation is obtained by evaluating the action on the fixed points of a variation as a
function of the endpoints of the variation. The momentum eigenstates can be
constructed in exactly this way. To see this, consider the non-relativistic free particle action
\beq
S=\int \frac{1}{2}m \dot{x}^{2}\ dt=\int p \cdot \dot{x} -\frac{p^{2}}{2m}\ dt.
\eeq
Hamilton's equations of motion obtained by finding the fixed points of the action are
\beqa
\dot{x}=p/m & & \dot{p}=0
\eeqa
whose general solution is $p=p_{0}=constant $, $x=x_{0}+p_{0}/m\  t$. Inserting this into the action we obtain
\beqa
S_{0}&=&\int^{t}_{t=0}\left(p_{0}\cdot\frac{dx}{dt}-\frac{p_{0}^{2}}{2m}\right) dt\nn\\
&=& \int^{x}_{x=0} p_{0} \cdot dx-\int^{t}_{t=0}\frac{p_{0}^{2}}{2m}dt \nn\\
&=& p_{0} \cdot x-\frac{p_{0}^{2}}{2m}t
\eeqa
The corresponding zeroeth order WKB state is clearly the momentum eigenstate corresponding to $p_{0}$. At this level,
it is only an approximate state, however, inserting it into the Schr\"{o}dinger equation shows that it is an exact
solution. Thus, in a sense it is an exact quantum state that is as close to classical as a a quantum state can be.
Because of this, it inherits many familiar properties of the classical solution. We note that the simplicity of the
above derivation, followed from the fact that the action evaluated on this class of solutions to Hamilton's equations
was a total derivative, and therefore independent of the path connecting the endpoints--one only needs to specify the
boundary data to evaluate the integral. We will see that the action for vacuum quantum gravity has the same property
which allows for exact WKB states. 

To illustrate the WKB nature of the Kodama states, we begin with the Holst action 
\beq
\frac{1}{k}\int \star e\w e\w R +\frac{1}{\beta}e\w e\w R -\frac{\lambda}{6}\star e\w e\w e\w e 
\eeq
whose equations of motion are
\beqa
&\star e\w  R -\frac{1}{\beta}e\w R -\frac{\lambda}{3}\star e\w e\w e=0& \\
&D\left(\star e\w e +\frac{1}{\beta} e\w e\right) =0.&
\eeqa
From the definition of the Weyl tensor $C=\frac{1}{4}\gamma^{I}\gamma^{J}C_{IJ\mu\nu}\frac{1}{2}dx^{\mu}\w dx^{\nu}$,
and the torsion $T=De$, one can show that the general solution to the above is of the form
\beqa
& R=\frac{\lambda}{3}e\w e +C &\\
&T=0. \label{GeneralSolution}&
\eeqa
We note that for $C=0$ the solution to the above is de Sitter space. As
in the non-relativistic case, we expect that to lowest order in $\hbar$, the wave function has the form
$\Psi=\mathcal{N}e^{i S_{0}/\hbar}$ where $S_{0}$ is the action evaluated on a particular solution to the equations of
motion.
Thus, we need to evaluate the action by inserting (\ref{GeneralSolution}) back into the Holst action and choosing a set
of boundary data which restricts one to a particular solution. In the following we will assume that the Weyl tensor is
small, keeping only first order terms in $C$, so we are dealing with small vacuum perturbations to
de Sitter space such as linearized gravitational waves propagating through an expanding universe. Setting the torsion
to zero in the Holst action annihilates the term involving the Immirzi parameter since $ e\w e \w R\sim e\w
DT$. One can view the Immirzi term as a control on the width of fluctuations of torsion in the path
integral\cite{Starodubtsev:MMgravity}. 
This term contains valuable information in the quantum theory so we will keep it by finding an
action which is equivalent to the Holst action at the fixed points. Inserting the equations of motion back into the
Holst action and dropping all $C\w C$ terms, we find
that on shell, the Holst action is equivalent to the topological action:
\beq
S_{0} \simeq \frac{3}{2k\lambda}\int \star R\w R +\frac{1}{\beta}R\w R.
\eeq
The first term is the Euler class and the second term is the second Chern class. Thus, as in the nonrelativistic case,
the action evaluated on the equations of motion is a total derivative. Since the above topological terms are tailored
to detect topological changes in the evolving manifold, it may be necessary to sum over past topological histories as
well as field configurations in the sum over histories---if the topological history of the manifold is trivial the above
terms are zero. Thus, for definiteness one might, for example, choose Hartle and Hawking's ``no boundaries" model and
sum over all closed topologies with a future spacelike boundary $\Sigma$ of a given spatial topology. With such a
choice, the action becomes
\beq
S_{0}\simeq \frac{3}{2k\lambda}\int_{\Sigma}\left(\star+\frac{1}{\beta}\right)\omega\w d\omega +\frac{2}{3}\omega\w
\omega
\w\omega.
\eeq
To make contact with the canonical theory, we partially fix the gauge to the time gauge, thereby reducing the gauge
group to $SU(2)$. In addition, recalling that the vanishing of the three-torsion emerged as a second class constraint
which was solved prior to canonical quantization, we set the three torsion to zero on the spacelike boundary $\Sigma$. We
proceed to rewrite the action in terms of the Ashtekar-Barbero connection, $A$, and a particular spatial triad
configuration whose Levi-Civita curvature, $R=R[E]$, we have seen will serve as a parameterization of a class of
states. The result is
\begin{equation}
S_{0}\simeq\frac{-3}{4k\Lambda \beta^{3}}\int_{\Sigma}Y_{CS}[A]-(1+\beta^{2})Y_{CS}[\Gamma] 
+2\beta(1+\beta^{2})Tr(K\wedge R_{\Gamma}), 
\end{equation}
which is precisely the argument of the generalized Kodama state $\Psi_{R}$ for a particular configuration of $R$. Thus,
we have shown that the generalized states are WKB states corresponding to first order perturbations about de Sittter
spacetime.

\section{The de Sitter state}
In the previous section, we suggested that the generalized states can be interpreted as WKB states corresponding to
first order vacuum perturbations about de Sitter space. Which state, then, is the WKB state corresponding to de Sitter
space itself? We recall that de Sitter space has multiple slicings in which the spatial topology is one of $\mathbb{R}^{3}$,
$\mathbb{S}^{3}$, or $\mathbb{H}^{3}$. We will focus on the $\mathbb{R}^{3}$ slicing where the metric takes the following
form:
\beq
ds^{2}=-dt^{2}+e^{2t/r_{0}}(dx^{2}+dy^{2}+dz^{2}).
\eeq
These coordinates do not cover the whole of de Sitter spacetime due to the presence of a cosmological horizon in the
expanding universe. It is of particular interest for our problem that in these coordinates, the 3-curvature is
identically zero, $^{(3)}R^{ij}_{ab}=0$ so the 3-space is flat Euclidean space. Thus, to pick out the de Sitter state,
we restrict the spatial topology to be $\mathbb{R}^{3}$ and turn the flat space condition into a quantum operator
equation
\beq
\int_{\mathbb{R}^{3}}\alpha\w\hat{R}\ |\Psi\rangle=0
\eeq
for all values of the test function, $\alpha$. The de Sitter solution is therefore the zero three-curvature
generalized Kodama state in $\mathbb{R}^{3}$ topology
\beqa
&|\Psi_{dS}\rangle =|\Psi_{R=0}\rangle &\\
&\langle A|\Psi_{R=0}\rangle = 
\mathcal{P}\exp\left[-\frac{3i}{4k\lambda\beta^{3}}\int_{\mathbb{R}^{3}} Y_{CS}[A]\right].&
\eeqa
Amongst all of the generalized states, this state is unique in being identically gauge and diffeomorphism invariant.
Furthermore, the loop transform of the above is well defined. Let $\Gamma$ be a framed graph whose edges and vertices
are labelled by representations of the quantum deformed group $SU_{q}[2]$ with deformation parameter
$q=e^{\frac{2\pi i}{\kappa+2}}$ where the level is given by $\kappa=\frac{3}{2G\lambda\beta^{3}}$.
The loop transform of the above is the Kauffman bracket of $\Gamma$:
\beq
\langle \Gamma |\Psi_{R=0}\rangle=K_{\Gamma}(q).
\eeq
We note the since the state is pure phase, the level of the Chern-Simons theory and the gauge group are real, so the
loop transform is more rigorous than that of the original Kodama state whose loop transform is complicated by the
complexification necessary in its construction. 

\subsection{Evidence of a cosmological horizon in the de Sitter solution}
There is another interesting consequence of the pure phase nature of
the state---we see evidence of a horizon. Since the state is pure phase, the deformation parameter is also pure phase .
At roots of unity when the level of the Chern-Simons theory takes integer values, the state is identically invariant
under large gauge transformations, and the representations of the corresponding quantum group terminate after a certain
level (see e.g. \cite{Kauffman:TemperleyLieb}). We note that the deformation parameter depends on three coupling
constants 
$k=8\pi G$, $\lambda$, and $\beta$. Since $G\lambda$ is observed to be very
small, $1/G\lambda$ is very large. Thus, one has freedom to fine tune the three coupling constants within
observational error to make the level, $\kappa$, an integer. We recall that in Loop Quantum Gravity, the area operator
has eigenvalues $A=8\pi G \beta\sqrt{j(j+1)}$ where $j$ is the representation of the edge piercing the operator valued
2-surface. A q-deformed spin network has edges labelled by representations of the q-deformed group. The deformation
parameter is given by $q=e^{\frac{2\pi i}{\kappa+2}}$ where $\kappa=\frac{3}{2 G \lambda \beta^{3}}$ is the level of
the
Chern Simons theory in point. If the deformation parameter is a root of unity the representations terminate at a maximum
spin $j_{max}$ where $j_{max}=\frac{\kappa}{2}$. At
large values of $\kappa$, and therefore $j_{max}$, we have $A\simeq 8\pi G \beta j$. This yields a maximum value for
the area of an indivisible surface:
\beq
A_{max}\simeq 2\pi \left( \frac{r_{0}}{\beta}\right)^{2}.
\eeq
This is on the order of area of the de Sitter horizon $A_{dS}=4\pi r_{0}^{2}$. We interpret this result as evidence of
the existence of a cosmological horizon in the quantum theory. It is of interest to note that in this simple model, we
see a new quantum mechanical feature---the Immirzi parameter, which has no classical effect in vacuum, is not only
significant at very small length scales in the quantum theory where it determines the scale on which Planck scale
discreteness occurs, but it also appears to play a significant role at extremely large, cosmological distances where
it modulates the de Sitter radius. However, a quantum cosmological horizon is not necessarily composed of
minimally divisible surfaces, and it is not known if this property holds in a more robust treatment.

\section{CPT Invariance}
In this section we will discuss the action of the discrete $C$, $P$, and $T$ operations on the generalized Kodama
states, showing that the wave functions violate $CP$ and $T$, while preserving $CPT$ symmetry. Each of the symmetries 
will have action both in the fibre and the base manifold. It will be useful to work in the Clifford algebra
representation to demonstrate the action of the symmetries in the fibre. We recall in the Clifford algebra, there are
two natural inner products. If $A$ and $B$ are arbitrary elements of the Clifford algebra, we have the standard inner
product given by $\langle A, B\rangle=Tr(AB)$, and the metric or $\star$-inner product given by $\langle
A,B\rangle_{\star}=Tr(\star AB)$. The generalized Kodama states utilize both inner products as the states can be
written in the following form:
\beq
\Psi^{\beta}_{R}[A]=\exp \left[-\frac{3i}{2k\Lambda}\int_{\Sigma}Tr(\star Y[\omega]+\beta^{-1} Y[\omega])\right] 
\label{CPTKodama}
\eeq
where $Y[\omega]=\omega \wedge d\omega +\frac{2}{3}\omega \wedge \omega \wedge \omega$ (no trace here) and $\omega$ is
the spin connection. In the above it is understood that the three torsion is set to zero on $\Sigma$ and a frame field
is fixed such that $R=R[E]$. We will consider each symmetry separately.

\subsection{Parity reversal}
On the base manifold, the action of parity simply inverts volume forms on the three-space. Since $Tr(Y)$ and
$Tr(\star Y)$ are ordinary three forms, they are inverted by parity. In addition, parity has action in the fibre which
can be deduced from the ordinary Dirac equation: $(i\gamma^{\mu}\partial_{\mu}-m)\psi[\vec{x},t]=0$. Under parity,
$x=(t,\vec{x})\rightarrow x'=(t,-\vec{x})$. A simple calculation shows that $\psi'(x')= P\psi(x)$, where $P=\eta
i\gamma^{0}$ and $\eta$ is an arbitrary phase factor, 
satisfies the space inverted Dirac equation $(i\gamma^{\mu}\partial '_{\mu}-m)\psi'(x')=0$. The Clifford algebra must
then transform in the adjoint representation so that
\beqa
&\mathcal{P}(\gamma^{\mu})=P\gamma^{\mu}P^{-1} & \nn\\
& (\gamma^{0},\gamma^{i})\rightarrow (\gamma^{0}, -\gamma^{i}) &
\eeqa
as expected. Under this transformation parity preserves the ordinary inner product on the Clifford algebra,
\beqa
& Tr(P AB P^{-1})=Tr(AB) & \nn\\
& \mathcal{P}(\langle A, B\rangle) =\langle A, B\rangle &
\eeqa
but inverts the metric inner product,
\beqa
& Tr(\star P AB P^{-1})=Tr(P^{-1}\star PAB)=-\langle A,B\rangle_{\star} &\nn\\
&\mathcal{P}(\langle A,B\rangle_{\star})=-\langle A,B\rangle_{\star}. &
\eeqa
The net effect on the wave functions \ref{CPTKodama} is an inversion of the Immirzi parameter:
\beq
\Psi^{\beta}_{R}\rightarrow \mathcal{P}(\Psi^{\beta}_{R}) =\Psi^{-\beta}_{R}.
\eeq
This is consistent with the general maxim with a growing body of evidence\cite{Soo:CPT, Rovelli:Torsion,
Freidel:Torsion, Randono:Torsion, Mercuri:Torsion},
\begin{quote}
\textit{The Immirzi parameter is a measure of parity violation built into the framework of quantum gravity.}
\end{quote}

\subsection{Time reversal}
Time reversal in a diffeomorphism invariant theory is somewhat subtle due to the effective dissapearance of time in
the canonical formalism. However, one should expect from general arguments that the quantum mechanical time reversal
operator should be anti-unitary. To see this, let us suppose we have an inner product which annihilates the
Hamiltonian constraint (the matrix elements of the scalar constraints are all zero) but preserves a causal ordering of
the ``in" and ``out" states. Consider then the inner product of $\Psi$ at $t_{2}$ and $\Phi$ at $t_{1}$ where
$t_{2}> t_{1}$ given by
\beq
\langle \Psi, t_{2}|\Phi, t_{1}\rangle .
\eeq
Since the inner product annihilates the Hamiltonian constraint, which generates time reparametrization, the above
inner product cannot depend on the particular values $t_{1}$ and $t_{2}$, though it does depend on their causal
ordering. Therefore, time reversal is equivalent to interchanging $t_{1}\leftrightarrow t_{2}$ (ignoring internal
degrees of freedom) so
the inner product
becomes $\langle \Phi, t_{2}|\Psi, t_{1}\rangle$. Again using the fact that the inner product only depends on the
causal ordering we conclude:
\beq
\mathcal{T}(\langle \Psi |\Phi \rangle)=\langle \Phi |\Psi \rangle =\langle \Psi |\Phi \rangle^{*}.
\eeq
Using the such an inner product to construct the connection representation, $\Psi[A]=\langle A| \Psi\rangle$, we
deduce the (partial) action of the time reversal operator is antiunitary:
\beq
\mathcal{T}(\Psi[A])=U \Psi^{*}[A]
\eeq
where $U$ is a unitary operator which represents the action of time reversal on any remaining internal degrees of
freedom.

To see the net effect of time reversal, we again appeal to the ordinary Dirac equation which we write in Hamiltonian
form: $i\partial_{t}\psi=H\psi$, where $H=i\gamma^{0}\gamma^{i}\partial_{i}-\gamma^{0}m$. The time reversed Dirac
equation is then $Ti\partial_{t}T^{-1}\psi'=THT^{-1}\psi'$, where $\psi'(\vec{x}, -t)=T\psi(\vec{x},t)$.
Time reversal must commute with the Hamiltonian so we must have
\beqa
&T\gamma^{0}T^{-1}=\gamma^{0}& \\\
& T\gamma^{i}T^{-1}=-\gamma^{i}&
\eeqa
and it must reverse the direction of time in the Dirac equation so 
\beq
TiT^{-1}=-i
\eeq
as expected from the previous arguments. On the inner products, we then have
\beqa
\langle A,B\rangle &\rightarrow& \langle A,B\rangle \\
 \langle A,B\rangle_{\star} &\rightarrow&  -\langle A,B\rangle_{\star}.
\eeqa
Thus, in total, the net effect on the generalized Kodama states is
\beq
\Psi^{\beta}_{R}\longrightarrow \mathcal{T}(\Psi^{\beta}_{R})=\Psi^{-\beta}_{R}.
\eeq
Evidently, time reversal undoes the action of parity inversion.

\subsection{Charge Conjugation}
On the surface the action of charge conjugation is simple: gravitons are there own antiparticles so charge conjugation
should not effect the wave-function of pure gravity. However, although one may be able to make this statement precise
in a perturbative context, uncovering the graviton from a non-perturbative framework is a difficult ordeal, and,
regardless, there may be subtle non-perturbative effects which determine the true action of the charge conjugation
operator. 

Like the time reversal operator, the charge conjugation operator is also anti-unitary. Specifically, on a Dirac
spinor, the charge conjugation operator take the form $\mathcal{C}(\psi)=\mathbb{C}\psi^{*}$ where
\beq
\mathbb{C}^{-1}\gamma^{I}\mathbb{C}=-\gamma^{I*}\ .
\eeq 
We expect the the charge conjugation operator to have a similar action on a quantum gravity wave function. However,
since the Kodama states are pure phase, they are not invariant under complex conjugation, and it can be easily checked
that the remaining action of $\mathbb{C}$ on the fibre indices has no effect. Thus, there appears to be a problem. The
resolution comes from the identification of $\int_{\Sigma}Y$ and $\int_{\Sigma} \star Y$ as topological charges. To
justify this we appeal to the gravitational conformal anomaly which states
\beqa
& d\ast J_{5} \sim R\wedge R &\nn\\
& \Delta Q_{5} \sim \Delta\int_{\Sigma}Y &
\eeqa
 where $J_{5}^{\mu}=\bar{\psi}\gamma_{5}\gamma^{\mu}\psi$ is the axial current and $Q_{5}$ is the axial charge which
is inverted by the charge conjugation operator.
Since the above equations are separately $P$ and $T$ invariant, in order for the them to be $CPT$ invariant, it must
be that the right hand side inverts under charge conjugation. Thus, if we identify it with a topological charge, the
charge conjugation must appropriately modify the topological structure on the bundle over $M$ bounded by $\Sigma_{1}\cup
\Sigma_{2}$ 
to invert $\int_{\Sigma}Y$. With this identification the charge operator acts as the identity on the generalized
Kodama states:
\beq
\mathcal{C}(\Psi_{R})=\Psi_{R}.
\eeq

\subsection{Net Effect}
In total we have shown that the states are CPT invariant:
\beqa
& \begin{array}{ccccccc}
  & \mathcal{C} &   & \mathcal{P} &  & \mathcal{T} & \nn\\
\Psi^{\beta}_{R}&\longrightarrow & \Psi^{\beta}_{R} &\longrightarrow &\Psi^{-\beta}_{R} &
\longrightarrow & \Psi^{\beta}_{R} 
\end{array} & \nn\\
& & \nn\\
&\mathcal{CPT}(\Psi^{\beta}_{R}) \longrightarrow \Psi^{\beta}_{R}\ .&
\eeqa
Finally, we note that although it is not known whether the states represent positive semi-definite energy states, one
cannot appeal to the standard argument from the analagous Chern-Simons state in Yang Mills theory to argue for negative
energies. The standard argument states that if there were a positive energy sector, the $CPT$ inverted sector, which
is also in the kernel of the constraints, must be a negative energy sector. Since $CPT$ does not invert the
generalized Kodama states, this argument no longer applies.

\section{The physical inner product and the Macdowell-Mansouri formulation of gravity}
In this section we discuss an interesting connection between the true, physical inner product defined by path integral
methods and the Macdowell-Mansouri formulation of gravity\cite{MMoriginal}. The kinematical inner product between two
states
of different 3-curvature given by, $\langle R'| R\rangle \sim \delta(\mathcal{R'}-\mathcal{R})$, is unlikely to be the
proper physical inner product, which is generally defined by a sum over histories as in the Hawking path integral and spin
foam methods. To this end we can formally write the true inner product as a sum over histories connecting two 3-curvature
states on the spacelike boundaries. That is, we take the boundary of our 4-dimensional manifold, M, to be two
3-dimensional spacelike hypersurfaces, $\Sigma_{2}$ and $\Sigma_{1}$, on which the the states $\Psi_{R'}$ and $\Psi_{R}$
are respectively defined. The expected physical inner product is then:
\beq
\langle \Psi_{R'}|\Psi_{R}\rangle_{phys}=\langle \Psi_{R'}|\int^{E_{2}}_{E_{1}}\mathcal{D}\omega 
\mathcal{D}e\ e^{iS_{EC+\beta}}|\Psi_{R}\rangle_{kin}
\eeq
where in the sum over histories we have fixed the spatial triad configurations $E_{2}$ and $E_{1}$ whose Levi-Civita
curvatures are $R'$ and $R$ respectively. In computing the path integral it will be useful to work in the connection
representation so that the total inner product takes the form
\beqa
\langle \Psi_{R'}|\Psi_{R}\rangle_{phys} &=& \int \mathcal{D}A'\mathcal{D}A\ \Psi^{*}_{R'}[A'] \Psi_{R}[A]
\int^{E_{2}}_{E_{1}}\mathcal{D}\omega \mathcal{D}e \ e^{iS_{H+\lambda}}\nn \\
&=& \int^{E_{2}}_{E_{1}}\mathcal{D}\omega \mathcal{D}e\  e^{-i\frac{3}{2k\lambda\beta}
\int_{M}R\wedge R -\frac{\beta}{2}\star R\wedge R}
e^{iS_{EC+\beta}}
\eeqa
where in the last line we have used the relation
\beq
e^{-i\frac{3}{2k\lambda\beta}\int_{M}R\wedge R-\frac{\beta}{2}\star R\wedge R}=\Psi^{*}_{R'}[A']\Psi_{R}[A].
\label{Psi*Psi}
\eeq
We now claim that the topological terms which enter into the physical inner product are precisely the topological
terms which enter into the Macdowell Mansouri action. To show this, we first need to extend the Macdowell Mansouri
formulation to include the Immirzi parameter. This has been accomplished in the context of $BF$ theory in
\cite{Smolin:MMaction, Starodubtsev:MMgravity}, however some of the topological terms in the resultant action, in
particular the Nieh-Yan
class, are relics of the $BF$ formulation and do not enter in the minimal prescription that we will present in the next
section.

\subsection{Adding the Immirzi Parameter}
We now need to extend the Macdowell-Mansouri action to include the Immirzi parameter. We refer the reader to the
appendix for
preliminaries on the Clifford algebra representation of the de Sitter group (section \ref{Appendix1}), and the
Macdowell-Mansouri action in our formalism (section \ref{Appendix2}). To this end, we first recall
how we add the Immirzi parameter to the Einstein Cartan action to give the Holst action. Beginning with the Einstein
Cartan action, 
\beq
S_{EC}=\frac{1}{k}\int_{M} \star e\wedge e\wedge R
\eeq
we simply perturb the curvature by its dual:
\beq
R\ \longrightarrow \ R-\frac{1}{\beta}\star R
\eeq
and the action becomes the Holst action
\beq
S_{H}=\frac{1}{k}\int_{M} \star e\wedge e\wedge R+\frac{1}{\beta}e\wedge e\wedge R.
\eeq
We will try using same trick on the Macdowell-Mansouri action. In our notation, the Macdowell-Mansouri action is given
by (see (\ref{Appendix2}) for details):
\beq
S_{MM}=-\frac{3}{2k\lambda}\int_{M}\star F\w F.
\eeq
The trick is to perturb the de Sitter curvature by its ``dual", 
\beq
F\ \longrightarrow \ F-\theta\star F,
\eeq
and possibly make appropriate adjustments to the coupling constant in order to regain the Holst action up to topological
terms. With the above substitution the Macdowell-Mansouri action becomes
\beqa
S_{MM+\beta}&=&\alpha\int_{M}\star (F-\theta\star F)\wedge (F-\theta\star F)\nn\\
&=& \alpha \int_{M} (1-\theta^{2})\star F\wedge F -\theta (\star F\wedge\star F-F\wedge F).
\eeqa
Making the identifications 
\beqa
\alpha(1-\theta^{2})=-\frac{3}{2k\lambda} & & \frac{2\theta}{1-\theta^{2}}=\frac{1}{\beta}
\eeqa
the action becomes 
\beq
S_{MM+\beta}=S_{topo}+S_{H+\Lambda}
\eeq
where $S_{H+\Lambda}$ is the Holst action with a positive cosmological constant and
\beq
S_{topo}=-\frac{3}{2k\lambda}\int_{M} \star R\wedge R+\frac{1}{\beta}R\wedge R.
\eeq
But this is precisely the negative of the argument of the generalized Kodama states in four-dimensional form!
Furthermore, this is exactly the term which enters into the true inner product as seen by equation (\ref{Psi*Psi}).
\beqa
\langle \Psi_{R'}|\Psi_{R}\rangle_{phys} &=& \int \mathcal{D}A'\mathcal{D}A\ \Psi^{*}_{R'}[A'] \Psi_{R}[A]
\int^{E_{2}}_{E_{1}}\mathcal{D}\omega \mathcal{D}e \ e^{iS_{H+\lambda}} \nn\\
&=& \int^{E_{2}}_{E_{1}}\mathcal{D}\Lambda \ e^{iS_{MM+\beta}}.
\eeqa

We conclude that the difference between the Macdowell-Mansouri formulation of gravity and the Einstein-Cartan
formulation is that the former already has the generalized Kodama states built into the theory as ground states. This
is similar to the two formulations of the $\theta$-ambiguity of Yang-Mills theory (see e.g. \cite{Ashtekar:book}). There
one finds that different
sectors of the phase space are connected via large gauge transformations. This ambiguity is reflected in the states,
which are not invariant under large gauge transformations but transform by a phase factor: $\Psi\rightarrow e^{i n
\theta}\Psi$. In an
attempt to salvage gauge invariance, one might normalize all the states by multiplying all states by the phase factor 
$e^{-i\frac{\theta}{8\pi^{2}}\int Y_{CS}}$ so that the $\theta$-ambiguity
is cancelled. However, the ambiguity simply reemerges in the inner product as the measure transforms to include a
factor of the second Chern class $e^{i\frac{\theta}{8\pi^{2}} \int F\wedge F}$. A similar phenomenon appears to 
be happening in the present situation.
\section{Concluding Remarks}
In the first paper of this two part series we showed that one can generalize the Kodama state to real values of the
Immirzi parameter and that the answer is surprising in that it opens up a large sector of physical states. Here we
showed that the sector not only includes the WKB state corresponding
to de-Sitter spacetime, but also includes WKB states corresponding to first order perturbations to de-Sitter spacetime
and possibly beyond. The states are directly related to two known topological field theories: the second Chern class
and the Euler characteristic. We isolated the de Sitter state by restricting to $\mathbb{R}^{3}$ topology and
demanding that the three-curvature vanishes. Among all the states, this state is unique in being identically gauge and
diffeomorphism invariant. Since the state is a pure phase whose argument is the Chern-Simons invariant, the q-deformed
spin-network transform is the Kauffman bracket. When the deformation parameter is a root of unity the $SU_{q}(2)$ representations
labeling the edges terminates at a finite value, thereby yielding evidence of a cosmological horizon. The full set of
states violate $CP$ while preserving $CPT$. Thus, although it is not known if the energy of the sector is positive
semi-definite, one cannot appeal to the standard argument for negative energies. Finally, we suggested that the
true physical inner product is likely to be a path integral connecting two states on the boundaries. Defined as such,
the inner product of two generalized Kodama states $\Psi_{R}$ and $\Psi_{R'}$ is equivalent to the path integral of the
Macdowell-Mansouri action with $R$ and $R'$ fixed on the end-caps. This behavior is reminiscent of the two
formulations of the $CPT$ problem in Yang-Mills theory.

\section*{Acknowledgments}
I would like to thank Lee Smolin and the Perimeter Institute for hosting me over the summer when some of the 
work in this paper was done. I would also like to thank Stephon Alexander for helping me understand the $CPT$
properties of the state and the relation to the gravitational conformal anomaly, and Don Witt for pointing out some
topological issues with the states.

\appendix
\section{Appendix}
\subsection{The Clifford representation of the de Sitter group}
\label{Appendix1}
In the section we discuss some preliminaries on the Clifford representation of the de Sitter group. 
For our purposes, the four-dimensional Lorentzian Clifford algebra will serve as a convenient basis for the Lie algebra
of the (anti)de-Sitter group. The Clifford algebra over a four dimensional Lorentzian $(-,+,+,+)$
signature metric is a sixteen dimensional algebra defined by the relation:
\beq
\gamma^{I}\gamma^{J}+\gamma^{J}\gamma^{I}=2\eta^{IJ}.
\eeq
When viewed as a complex vector space, the Clifford algebra forms a sixteen (complex) dimensional vector space with a
convenient basis given by
\beqa
&1&\nn\\
&\frac{1}{2}\gamma^{I}&\nn\\
&\frac{1}{4}\gamma^{[I}\gamma^{J]}&\nn\\
&\frac{1}{2}\gamma^{I}\star&\nn\\
&\star&
\eeqa
where $\star=-i\gamma_{5}=\gamma^{0}\gamma^{1}\gamma^{2}\gamma^{3}$ and the numerical factors are inserted for later
convenience. When the algebra is viewed as a complex Lie algebra, it is isomorphic to $gl(4,\mathbb{C})$.
We first consider the group that preserves the Dirac inner product
\begin{equation}
i\overline{\Phi}\Psi=i\Phi^{\dagger}\gamma^{0}\Psi
\end{equation}
The group which preserves this inner product must satisfy $\gamma^{0}g^{\dagger}\gamma^{0}=-g^{-1}$ which translates
into the Lie algebra condition:
\beq
\gamma^{0}\lambda^{\dagger}\gamma^{0}=\lambda.
\eeq
The semi-simple Lie algebra which satisfies this relation is spanned by the basis 
$\{\frac{1}{4}\gamma^{[I}\gamma^{J]}, \frac{1}{2}\gamma^{I}\gamma_{5}, \frac{1}{2}\gamma^{I}, \gamma_{5} \}$. This Lie
algebra is a basis for the four dimensional conformal algebra $su(2,2)$, which can easily be seen by
choosing a basis in which $i\gamma^{0}= diag(1,1,-1,-1)$. The de Sitter and anti-de Sitter algebras are then spanned by
the {\it real} subalgebras $\{\frac{1}{4}\gamma^{[I}\gamma^{J]}, \frac{1}{2}\gamma^{I}\gamma_{5} \}$ for de Sitter and
$\{\frac{1}{4}\gamma^{[I}\gamma^{J]}, \frac{1}{2}\gamma^{I} \}$ for anti-de Sitter. Restricting now to the de Sitter
group, we can construct a de Sitter connection with connection coefficients $\Lambda=\omega+\frac{1}{r_{0}}e \gamma^{5}$.
When the gauge group breaks down to $SO(3,1)$, we identify $\omega=\omega_{IJ}\frac{1}{4}\gamma^{[I}\gamma^{J]}$
with the spin connection, $e=e_{I}\frac{1}{2}\gamma^{I}$ with the four dimensional frame field, and
$r_{0}=\sqrt{\frac{3}{\lambda}}$
with
the de Sitter radius. To see this, we first consider the curvature of $\Lambda$ taking into consideration the above
identifications:
\beqa
F &=& d\Lambda+\Lambda\wedge \Lambda \\
&=& R+\frac{1}{r_{0}}T\gamma^{5}-\frac{\lambda}{3}e\wedge e.
\eeqa
where, $T=T_{I}\frac{1}{2}\gamma^{I}$ is the torsion and $R=R_{IJ}\frac{1}{4}\gamma^{I}\gamma^{J}$ is the $SO(3,1)$
curvature. Once the gauge group is broken to $SO(3,1)$, de Sitter spacetime is then given simply by the condition\footnote{Technically
this is not de Sitter
spacetime because the ``frame field", e, is not necessarily invertible. In this sense, the Macdowell-Mansouri
formulation is a natural extension of general relativity to include degenerate metrics.} $F=0$.

\subsection{The Macdowell-Mansouri Action}
\label{Appendix2}
We can now attempt to construct an action for general relativity using only the curvature $F$. The most natural action
to try is the Yang-Mills action $\int \ast F\wedge F$ where $\ast$ is the dual on the base manifold. However, when the
gauge group reduces
to $SO(3,1)$ we are left with an $SO(3,1)$ Yang-Mills term $\int \ast R\wedge R$ which is not in any gravity theory. We
can solve this problem by instead using the dual on the fibre indices given by $Tr(\star AB)$. This breaks de Sitter
invariance since $\star$ is not preserved by the de Sitter group, however, the symmetry is reduced naturally to
$SO(3,1)$ so it still could yield general relativity without being strictly de Sitter invariant. The
Macdowell-Mansouri action is then given by
\beqa
S_{MM} &=& \alpha'\int\star F\wedge F\nn\\
&=& \alpha'\int \star R\wedge R-2\frac{1}{r_{0}^{2}}\star e\wedge e\wedge R+\frac{1}{r_{0}^{4}}\star e\wedge e\wedge e\wedge
e.
\eeqa
With the identification $\alpha'=\frac{-3}{2k\lambda}$, the action is precisely the Einstein-Cartan action with a
positive cosmological constant plus the Euler class, 
\beq
\int \star R\wedge R=\int\frac{1}{4}\epsilon_{IJKL}R^{IJ}\wedge R^{KL}
\eeq
which, being topological, does not affect the local equations of motion.

\bibliography{GKSII}

\end{document}